\documentstyle[12pt,twoside,fleqn,espcrc1,epsf]{article}

\def\Lam{\Lambda}  \def\Sig{\Sigma}
\def\wave{\simeq}

\def\di{{\Delta I}}

\title{Direct Quark Mechanism for Weak $\Lam N \leftrightarrow
       NN$ Processes}
\author{Makoto Oka\address{Department of Physics,
        Tokyo Institute of Technology \\
        Meguro, Tokyo, 152 JAPAN}%
        \thanks{email: oka@th.phys.titech.ac.jp}}

\begin{document}

\maketitle

\date{\today}

\begin{abstract}
 Two body weak processes $\Lambda N \leftrightarrow NN$ are studied from 
 the viewpoint of quark substructure of baryons.
 They can be studied in nonmesonic weak decays of hypernuclei 
 and also hyperon production in the $NN$ scattering.
 The direct quark mechanism in contrast to the one-pion exchange mechanism 
 is shown to give sizable $\di=3/2$ contribution in the $J=0$ channel. 
 Nonmesonic decay rates of $^4_{\Lambda}$H, $^4_{\Lambda}$He, and  
 $^5_{\Lambda}$He agree with the current experimental 
 data fairly well.  Especially the $n/p$ ratio is enhanced by the 
 direct quark contribution.
 The cross sections for $pn\to \Lam p$ scattering are studied in the 
 quark cluster model approach.  
\end{abstract}

\section{Introduction}

Recent experimental and theoretical studies of weak decays of
hypernuclei have generated renewed interest on nonleptonic
weak interactions of hadrons.
A long standing problem is the dominance of $\di=1/2$ amplitudes,
called the $\di=1/2$ rule, in the strangeness changing transitions.
Nonleptonic decays of kaons and $\Lam$, $\Sig$ hyperons are dominated
by the $\di=1/2$ transition but it is not clear whether this dominance
is a general property of all nonleptonic weak interactions.
In fact, the weak effective interaction which is derived
from the standard model including the perturbative QCD
corrections contains a significantly large $\di=3/2$
component{\cite{hweak}}.
It is therefore believed that nonperturbative QCD corrections,
such as hadron structures and reaction mechanism are
responsible for suppression of $\di=3/2$, and/or enhancement
of $\di=1/2$ transition amplitudes.

From this viewpoint, decays of hyperons inside nuclear medium provide
us with a unique opportunity to study new types of nonleptonic
weak interaction, that is, two- (or multi-) baryon processes,
such as $\Lam N\to NN$, $\Sig N\to NN$, etc.
These transitions consist the main branch of hypernuclear
weak decays because the pionic decay $\Lam\to N\pi$ is suppressed
due to the Pauli exclusion principle for the produced nucleon.

A conventional picture of the two-baryon decay process,
$\Lam N\to NN$, is the one-pion exchange between the baryons,
where the $\Lam N\pi$ vertex is induced by the weak
interaction
{\cite{BD}}.
In $\Lam N\to NN$, the relative momentum of the final state nucleons
is about 400 MeV/c, much higher than the nuclear Fermi momentum.
The nucleon-nucleon interaction at this momentum is dominated
by the short-range repulsion due to heavy meson exchanges and/or
to quark exchanges between the nucleons.
It is therefore expected that the short-distance interactions
will contribute to the two-body weak decay as well.
Exchanges of $K$, $\rho$, $\omega$, $K^*$ mesons
and also correlated two pions in the nonmesonic
weak decays of hypernuclei have been
studied{\cite{mesonex,Ramos}} and it is found that the kaon
exchange is significant, while the other mesons contribute
less{\cite{Ramos}}.

Several studies have been made on effects of quark
substructure{\cite{CHK,MS,ITO}}.
In our recent analyses{\cite{ITO,IOMI}}, we employ an effective weak
hamiltonian for quarks,
which takes into account one-loop perturbative QCD corrections
to the $W$ exchange diagram in the standard model{\cite{hweak}}.
It was pointed out that the $\di=1/2$ part of the hamiltonian is
enhanced during downscaling of the renormalization point
in the renormalization group equation.
Yet a sizable $\di=3/2$ component remains in the low energy effective
weak hamiltonian.
We proposed to evaluate the effective hamiltonian in the six-quark wave
functions of the two baryon systems and derived the ``direct quark (DQ)''
weak transition potential for $\Lam N \to NN${\cite{ITO,IOMI}}.
Our analysis shows that the DQ contribution
largely improves the discrepancy between the meson-exchange theory
and experimental data for the ratio of the neutron- and proton-induced
decay rates of light hypernuclei.
It is also found that the $\di=3/2$ component of the effective hamiltonian
gives a sizable contribution to $J=0$ transition amplitudes.
Unfortunately, we cannot determine the $\di=3/2$ amplitudes
unambiguously from
the present experimental data{\cite{Schu}}.

\section{Direct Quark Transition Potentials}

The DQ transition  
takes place only when $\Lam$ overlaps with a
nucleon in hypernuclei and therefore predominantly in the relative
$S$-states of $\Lam N$ systems.
The two-body transition potentials for
the initial $\Lam N (L=0)$ and  the final $NN (L=0,1)$ states  are computed 
in the DQ mechanism and are compared with those in the one-pion
exchange (OPE) mechanism. 
Because of quark antisymmetrization effects, the transition
potential contains nonlocal components and as the transition may
break the parity invariance, it also contains derivative terms.
The general form of the transition potential is
\begin{eqnarray}
 V^{\ell\ell'}_{ss'J}(r,r') &=& \langle NN: \ell' s' J|V(\vec r',
  \vec r) |\Lam N : \ell s J\rangle \nonumber\\
  &=& V_{loc}(r)\, {\delta(r-r')\over r^{2}}
    + V_{der}(r)\, {\delta(r-r')\over r^{2}} \partial_{r}
    +V_{nonloc} (r',r)
\end{eqnarray}
Fig.\ 1(a) shows the local part of the DQ and OPE potentials
and Fig.\ 1(b) shows the nonlocal part
of the DQ potential for the $\Lam p (^1S_0)\to np (^1S_0)$ transition.
It can be seen in the local potential that the DQ is dominant at
short distances, $r<1.4$ fm, while the OPE in the region $r<2$ fm is
modified strongly by the form factor introduced according to \cite{MG}.
Both the local and nonlocal potentials show that
$\di=3/2$ contribution is significant.
Such strong violation of the $\di=1/2$ dominance
takes place in $J=0$ transitions,
$\Lam N (^1S_0)\to NN (^1S_0)$ and $\Lam N (^1S_0)\to NN (^3P_0)$.

\begin{figure}[htb]
\begin{minipage}[t]{80mm}
   \epsfysize = 80 mm
   \centerline{ \epsfbox{1s0loc.EPSF} }
\caption{The local part of the $\Lambda N\, ^{1}S_{0} \to NN\,^{1}S_{0}$ 
transition potential.}
\end{minipage}
\hspace{\fill}
\begin{minipage}[t]{65mm}
   \epsfysize = 70 mm
  \centerline{ \epsfbox{1s0nl.EPSF} }
\caption{Contour plot of the DQ nonlocal transition potential.}
\end{minipage}
\end{figure}

Because the OPE potential is determined phenomenologically, the relation
between DQ and OPE is not trivial.  
In order to check the consistency, we relate the $\pi N\Lam$
coupling constant to a baryon matrix element of the weak hamiltonian
for quarks by using the chiral property.
In the soft-pion limit, the soft-pion theorem leads to{\cite{soft-pion}}
\begin{equation}
\lim_{q\to 0}\langle\pi^{0}(q) n|H_{W}|\Lam\rangle = {i\over f_{\pi}}
         \langle n|[Q_{5}^{3},H_{W}]|\Lam\rangle
\end{equation}
where $Q_{5}^{3}$ is the third component of the axial charge operator.
The weak hamiltonian $H_{W}$ consists only of left-hand currents and
the flavor-singlet right-hand currents, and therefore satisfies
$[Q_{5}^{3},H_{W}] = - [I^{3},H_{W}]$, where $I^{3}$ is the third
component of the isospin operator.
Taking the parity conserving part, we obtain\cite{soft-pion}
\begin{equation}
\lim_{q\to 0}\langle\pi^{0}(q) n| H_{PV} |\Lam\rangle = {-i\over 2f_{\pi}}
         \langle n| H_{PC} |\Lam\rangle
         \label{eq:sp}
\end{equation}
The LHS of this equation is proportional to the effective $\Lam N\pi$
weak coupling constant, defined by
\begin{equation}
      H_{eff} = ig_{w}\bar\psi_{n} (1+\lambda\gamma^{5})\pi^{0}\psi_{\Lam}
\end{equation}
where the phenomenologically determined value of the coupling
constant is $|g_{w}|=2.3\times 10^{-7}$ (dimensionless).
The RHS of eq.(\ref{eq:sp}) can be evaluated by using the present
quark model. Then using the Goldberger-Treiman relation we obtain
\begin{equation}
        g_{w} = g_{\pi NN} {\langle n| H_{PC} |\Lam\rangle\over 2M_{N}g_{A}}
        = -3.5\times 10^{-7}
\end{equation}
in our quark model.  Here $g_{\pi NN}$ ($=-13.26$) is defined for
\begin{equation}
                H_{\pi NN} = g_{\pi NN} \bar\psi_{N} i\vec\tau\cdot\vec\pi
                \gamma^{5}\psi_{N}
\end{equation}
This value is about 30\% larger than the phenomenological one.
The reason for the overestimation may reside in the soft-pion assumption
as well as in the approximation of
the baryon wave functions, that is, the orbital quark wave functions
of $\Lam$ and $n$ are assumed to have the same Gaussian form.
However, by this estimate we can determine the relative phase of the
phenomenological $\Lam N\pi$ coupling constant and the weak quark
hamiltonian.
Namely, the product of $g_{w}$ and $g_{\pi NN}$ in the OPE potential
must be positive in order to be consistent with the weak hamiltonian
we employ.

Note that the single baryon matrix element  $\langle n| H_{PC} |\Lam\rangle$
has no contribution from the $\di=3/2$ part of $H_{W}$ and
therefore eq.(\ref{eq:sp}) guarantees the $\di=1/2$ dominance.
In fact, the same argument leads to the $\di=1/2$ rule for the
pionic decays of hyperons as all the single baryon matrix elements,
such as $\langle p| H_{PC} |\Sigma^{+}\rangle$,
$\langle \Sigma^{0}| H_{PC} |\Xi^{0}\rangle$, have no $\di=3/2$
contribution if we assume a valence quark picture for the baryon
according to {\cite{Pati-Woo}}.
The soft-pion limit also predicts that the $\Sigma^{+}\to n\pi^{+}$
decay is not allowed because the transition from $\Sigma^{+}(I_{3}=+1)$
to $n(I_{3}=-1/2)$ is possible only through the $\di=3/2$ weak
hamiltonian.  Indeed, $\Sigma^{+}\to n\pi^{+}$ is strongly suppressed
experimentally.

\section{Nonmesonic Weak Decay of Light Hypernuclei}

In \cite{ITO,IOMI}, we calculated the nonmesonic decay rates of
$^{5}_{\Lam}$He, $^{4}_{\Lam}$He, and $^{4}_{\Lam}$H in the DQ and
OPE mechanisms.
The $S$-shell hypernuclei are most suitable for the study of the
microscopic mechanism of the weak decay as their wave functions are
relatively simple and contain only $\Lam N (L=0)$ states.
They also enable us to select spin-isospin components for the weak
decay.

We here use a realistic $\Lam$-nucleus wave function constructed from
the YN G-matrix interaction (YNG potential\cite{Yamamoto,Motoba})
with the $(0s)^{n}$ harmonic oscillator wave function for the nucleus.
We also include the short-range correlation, the Nijmegen D
correlation for OPE and the Gaussian soft cutoff correlation according
to the baryon's quark wave function for DQ.

\begin{table}[hbt]

\caption{ Calculated nonmesonic decay rates of $^5_{\Lambda}$He
          (in unit of $\Gamma_{\Lambda}$). 
          $\Gamma_{nm} = \Gamma_{p}+\Gamma_{n}$ and
          $R_{np} = \Gamma_{n}/\Gamma_{p}$. }
\begin{center}
\begin{tabular}{|lrl|c|c|c|}
\noalign{\hrule}
     & Channels & &  OPE  &  DQ &  OPE $+$ DQ
            \\
\noalign{\hrule}
  $p\Lambda \to pn$ & ${}^{1}S_0 \to {}^{1}S_0$ & $ a_{p}   $ 
    & 0.0002  & 0.0167 &  0.0188  \\
                    &           $\to {}^{3}P_0$ & $ b_{p}   $ 
    & 0.0031  & 0.0113 &  0.0026  \\
                    &  ${}^{3}S_1\to {}^{3}S_1$ & $ c_{p}   $
    & 0.1022  & 0.0548 &  0.2612  \\
                    &           $\to {}^{3}D_1$ & $ d_{p}   $ 
    & 0.0415  & 0      &  0.0415  \\
                    &           $\to {}^{1}P_1$ & $ e_{p}   $ 
    & 0.0346  & 0.0064 &  0.0207  \\
                    &           $\to {}^{3}P_1$ & $ f_{p}   $ 
    & 0.0093  & 0.0353 &  0.0763  \\
\noalign{\hrule}
  $n\Lambda\to nn$  &  ${}^{1}S_0\to {}^{1}S_0$ & $ a_{n}   $ 
    & 0.0003  & 0.0407 &  0.0356  \\
                    &           $\to {}^{3}P_0$ & $ b_{n}   $ 
    & 0.0063  & 0.0069 &  0.0264  \\
                    &  ${}^{3}S_1\to {}^{3}P_1$ & $ f_{n}   $ 
    & 0.0185  & 0.0648 &  0.1437  \\
\noalign{\hrule}
 && $\Gamma_{p}   $ & 0.191 & 0.125 & 0.421 \\
 && $\Gamma_{n}   $ & 0.025 & 0.112 & 0.206  \\
 && $\Gamma_{nm}  $ & 0.216 & 0.237 & 0.627  \\
 && $ R_{np}      $ & 0.132 & 0.903 & 0.489 \\
\noalign{\hrule}
\end{tabular}
\end{center}
\end{table}

We found that for $^{5}_{\Lam}$He, that the DQ amplitudes are
dominant over OPE in $J=0$ transitions ($a_{p}$, $b_{p}$, $a_{n}$)
as well as $^{3}S_{1}\to^{3}P_{1}$ ($f_{p}$, $f_{n}$) transitions
(Table 1).
Also large $\di=3/2$ contributions are predicted for $J=0$
transitions ($a_{p}$, $b_{p}$, $a_{n}$, $b_{n}$).
Accordingly, the neutron induced decay rate $\Gamma_{n}$ is enhanced
from 0.025 (OPE only) to 0.206 (OPE+DQ) and thus improves the $n/p$
ratio.  This is still short of the experimental value as $\Gamma_{p}$
is dominated by the $J=1$ amplitudes from the strong one-pion tensor
component.  It was suggested that the kaon exchange may be important
in reducing the $\Gamma_{p}(J=1)$ \cite{Ramos}.

For $^{4}_{\Lam}$H, because no $J=1$ $\Lam p\to pn $ contributes, the
DQ transitions become dominant.  Both $\Gamma_{p}$ and $\Gamma_{n}$
are significantly enhanced by DQ:
$\Gamma_{p}= 0.004$ (OPE) to $\Gamma_{p}= 0.047$ (DQ+OPE), and
$\Gamma_{n}= 0.017$ (OPE) to $\Gamma_{n}= 0.126$ (DQ+OPE),
and the total decay rates is predicted to be
$\Gamma_{nm} = 0.174$.
Although an indirect estimate \cite{Outa} suggests $\Gamma_{nm}
\wave 0.17\pm 0.11$, a direct measurement is very much anticipated.

For $^{4}_{\Lam}$He, the neutron induced decay is suppressed, which
seems consistent with experiment.

\begin{figure}[htb]
\begin{center}
   \epsfysize = 90 mm
 \epsfbox{cross.EPSF} 
\caption{The $NN\to \Lam N$ scattering cross sections.}
\end{center}
\end{figure}

\section{Weak $pn\to \Lam p$ Production}

The two-body weak interaction can be directly studied in the $\Lam$ 
production scattering: $pn\to \Lam p$.
The cross sections of such two-body processes were calculated in the 
meson exchange model\cite{weaksc}, which claim 
$10^{-15}\wave 10^{-16}$ barn.
The cross section is dominated by the $^{3}S_{1}$ and $^{3}D_{1}$
channels and shows significant parity violation ($\wave 25$\%).

We have calculated the $pn\to p\Lam $ cross section in the full 
six-quark cluster model with DQ+OPE weak transitions.
We employ the standard quark cluster model for the initial $NN$ and 
the final $\Lam N$ scattering wave functions\cite{QCM}.  
The model contains the 
nonrelativistic quark kinetic energy, one-gluon exchange interaction 
(central + spin-spin + spin-orbit parts), and the meson exchange 
potentials for the nonets of scalar, pseudoscalar and vector mesons.
The parameters in the meson exchange potential are tuned so as to 
reproduce the $NN$ scattering phase shifts up to $E_{cm}\approx 350$ 
MeV.  For the $\Lam N$ scattering,  the $\Sig N(I=1/2)$ channels are 
always coupled.

The transition T matrix is calculated in the distorted wave Born 
approximation (DWBA) formulation.  
Our preliminary results consider only the weak transition to $\Lam 
N$ (not $NN\to \Sig N$) and only the $^{1}S_{0}$ and $^{3}S_{1}$ 
$\Lam N$ final states.
We found that the cross sections are of order of $10^{-16}$ barn, and 
are sensitive to the choices of the form factor for OPE and 
parameters in the $\Lam N$ final state interactions.
Fig.\ 2 shows the (preliminary) results, which indicate that the OPE 
is dominated near threshold.  The amplitudes are dominated by the 
$J=1$ contribution.  We also found that the PV/PC ratio depends on 
the $\Lam N$ energy rather strongly.  It becomes as large as 60\% 
at $E_{cm}\approx 50$ MeV.
Inclusion of the $\Sig N$ channels and the higher partial waves 
have to be done further as well as study of dependencies on
the parameters in the final state interactions.
Further study is under way.

\section{Conclusion}

In conclusion, we would like to stress that nonmesonic weak 
decays of hypernuclei and weak hyperon productions are rich 
sources of interesting information on the nonleptonic weak 
interactions of hadrons.
Here we have discussed the possibility of the direct quark 
contributions.
We have found that the $\di=1/2$ rule may be significantly violated 
in hypernuclear weak decays, which can be tested in decays of the 
$S$-shell hypernuclei.
It is crucial to have a high quality measurement of the decays of
$^{4}_{\Lam}$H.

Recently, we pointed out that soft $\pi^{+}$ emissions in weak 
decays of hypernuclei probe the $\di=3/2$ part of the nonmesonic weak 
decay\cite{piplus}. 
We found that the soft $\pi^{+}$ can be emitted in the two-body 
process which is induced by the $\di=3/2$ weak interaction.
The DQ transition potential predicts significant $\pi^{+}$
emissions in light hypernuclei.
This opens a new possibility to check consistency in the weak decay 
mechanisms.

\bigbreak
The author thanks Drs. Takashi Inoue, Sachiko Takeuchi, Toshio Motoba
and Kazunori Itonaga for collaborations regarding this report.
This work is supported in part by the Grant-in-Aid for scientific
research (C)(2)08640356 and Priority Areas (Strangeness Nuclear
Physics) of the Ministry of Education, Science and Culture of Japan.

\def \vol(#1,#2,#3){{{\bf {#1}} (19{#2}) {#3}}}
\def \NP(#1,#2,#3){Nucl.\ Phys.\          \vol(#1,#2,#3)}
\def \PL(#1,#2,#3){Phys.\ Lett.\          \vol(#1,#2,#3)}
\def \PRL(#1,#2,#3){Phys.\ Rev.\ Lett.\   \vol(#1,#2,#3)}
\def \PRp(#1,#2,#3){Phys.\ Rep.\          \vol(#1,#2,#3)}
\def \PR(#1,#2,#3){Phys.\ Rev.\           \vol(#1,#2,#3)}
\def \PTP(#1,#2,#3){Prog.\ Theor.\ Phys.\ \vol(#1,#2,#3)}
\def \PTPsup(#1,#2,#3){Prog.\ Theor.\ Phys.\ Suppl.\ \vol(#1,#2,#3)}
\def \ibid(#1,#2,#3){{\it ibid.}\         \vol(#1,#2,#3)}
\def\MO{M.~Oka} \def\etal{{\it et al.}}

\end{document}